\documentstyle[12pt,epsf]{article}

\begin{document}

\thispagestyle{empty}
\normalsize

\begin{center}
{\Large \bf {
Are there non-strange low-lying penta-quarks and can we 
understand their width
}} \\[15mm]
{\large {R.W. Gothe$^{a}$ and S. Nussinov$^{a,b}$}} 
\\[5mm]
{\it {
$^a$Department of Physics \& Astronomy, University of South Carolina,
Columbia, USA\\
$^b$School of Physics \& Astronomy, University of Tel Aviv, Ramat Aviv 
Tel-Aviv, Israel
}} \\[15mm]
{\bf {Abstract}} \\[3mm]
\end{center}

\noindent
We argue that the lightest isospin 1/2 partners of the $Z^+(1530)$ 
$\bar{s}uudd$ penta-quark predicted by Diakonov, Petrov and Polyakov 
are not the $N(1710)$ mixed anti-decuplet states, but the pure 
non-strange $\bar{u}(ud)(ud)$ and $\bar{d}(ud)(ud)$ penta-quark 
states which may lie as low as $1200\,MeV$. The expected low width 
of a few $MeV$ of such a putative state may explain why it was missed 
in phase shift analyzes of pion-nucleon scattering.
\\[15mm]

\begin{figure}[hb!]
\centering
\vspace*{-0mm}
\mbox{\hspace*{0mm} \epsfxsize=140mm \epsffile{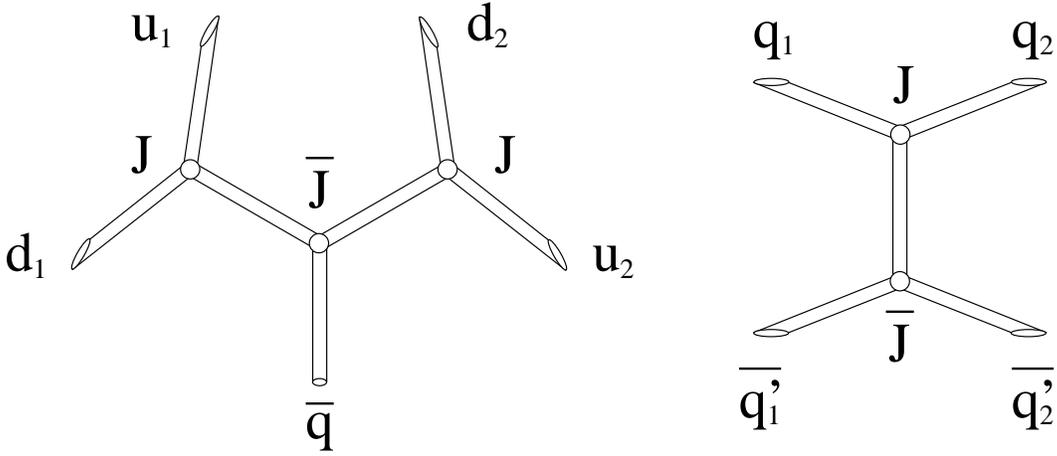}}
\vspace*{-0mm}
\caption{Schematic description of the net-works of the chromoelectric 
flux tubes in case of the tetra-quark (right) with one junction $J$ and one
anti-junction $\bar{J}$ and the penta-quark (left) with two junctions 
and one anti-junction. Three fluxes belonging in the fundamental 
${\bf{3}}$ color representation are flowing into a junction J and 
three such fluxes are flowing out of an anti-junction $\bar{J}$.
Any outgoing/incoming flux in color ${\bf{3}}$ is equivalent to an 
outgoing/incoming flux in color ${\bf{\bar{3}}}$ representation. 
The fluxes are locally coupled 
to a singlet via $\varepsilon_{abc} \Phi_a\Phi_b\Phi_c$ with 
$a,b,c$ the three colors. The drawing shows an idealized case where the 
length $a$ of the flux tubes joining $J$ and $\bar{J}$ are significantly 
larger than the radius $b$ of tube.
}
\label{pic.tube}
\end{figure}

\noindent
Exotic tetra-, penta- and hexa-quarks have been discussed in the frame-work 
of QCD and various approximations thereof for more than three decades. 
The generally accepted point of view was that such states which certainly 
should exist are numerous, broad and hence blend in a continues background. 
The anomalous lightness of the Nambu-Goldstone pions allows most exotics to 
decay into a stable baryon and pions or just into pions with large phase 
spaces in all cases. It was also argued that decay rates of non exotics 
as $q\bar{q}$ mesons or $qqq$ baryons requiring the creation of an 
extra and often even specific $q\bar{q}$ pair are suppressed \cite{Witten1}
relative to the "fall-apart" decay of exotics like the 
$\Theta^+ \equiv \bar{s}uudd \rightarrow \bar{s}u + udd \equiv K^+n$, 
where all final quarks are already present in the initial state.
\\
If pionic decays are partially blocked by the Zweig rule, suppressing 
$s\bar{s}$ annihilation into the lighter $u\bar{u}$ or $d\bar{d}$ 
quarks, the relevant states like the $I=0$ ~$f(980)$ and $I=1$ ~$a(980)$, 
which following Jaffe \cite{Jaffe1} we take as 
$s\bar{s}q\bar{q}$~\footnote{$q$ refers to the lightest quarks 
$(u,d)$, $Q$ to heavy quarks $(c,b)$ and the 
intermediate $s$ is mentioned explicitly.} configurations, 
can have reasonable $50-100\,MeV$ widths despite of the large phase 
space for pionic decays. It was then suggested that in certain
heavier quark systems, such as the $\bar{c}suud$ penta-quark 
\cite{Lipkin1,Zouzou,Fleck} or $ssuudd$ $\Lambda \Lambda$-hexa-quark 
\cite{Jaffe2} states, the favorable strong hyperfine 
interaction can generate stable multi-quark states which decay 
only weakly. It has been noted by several authors 
\cite{Tornqvist,Wise,Gelman}
that tetra-quarks, particularly the $cc\bar{q}\bar{q}$ or $c\bar{c}q\bar{q}$,  
are even more likely to be discovered. This was indeed verified in the 
Belle experiment \cite{Belle}, where a remarkable narrow peak was found 
in the $J/\Psi \pi^+ \pi^-$ channel. This state which we believe has 
the quantum numbers of a $D^{\ast} \bar{D}$ in $s$-wave, namely 
$J^{\pi} = 1^+$, forms  readily in the $B$ decay involving two charm quarks 
in close spatial proximity and with reasonably low relative momenta. Its 
annihilation into $J/\Psi$ and  pions is still suppressed by the need to 
have the $c\bar{c}$ in the $J/\Psi$ configuration causing its 
remarkable narrowness. This may be the first of a host of other 
$Q\bar{Q}q'\bar{q}$ exotica with $q',q \in \{u,d,s\}$. If the 
Babar $D_s(2317)$ state \cite{Barnes} is a 
$c\bar{s} \cdot (u\bar{u} + d\bar{d})/\sqrt{2}$ rather than $p$-wave 
$c\bar{s}$ state \cite{Cahn1}, QCD inequalities \cite{Nussinov1} suggest 
the existence of a third member in the series \cite{Nussinov2} starting 
with $f(980) \equiv s\bar{s}q\bar{q}$, 
namely the $I = 0$ ~$c\bar{c}q\bar{q}$ state with a mass $m \le 3670\,MeV$ 
and a narrow width decaying into $\eta_c \eta$.
\\
A $\bar{s}uudd$ penta-quark resonance at $m(\Theta^+) = 1540\,MeV$ has been 
seen in $K^+ n$ invariant mass distributions in various real photon 
experiments off deuterons \cite{Spring,Jlab} and  
protons \cite{ELSA} by analyzing the $K^- K^+ n$ and  
$K_{s}^{0} K^+ n$ final states without being swamped by multiple pion 
complex final states. This resonance has also been reported in $K^+$-Xenon 
scattering \cite{DIANA} and the possible existence of a very weak 
evidence in $K^+d$ cross section data in the PDG 
(Particle Data Group compilation) \cite{PDG} was noted \cite{Nussinov3}.
\\
The most remarkable feature of the new state is  the narrow width 
$\Gamma_{\Theta+} < 25\,MeV$  seen in $\gamma d$ and $\gamma p$ 
experiments where this upper bound is given by the experimental
resolution. A much more stringent bound of $\Gamma < 3-6\,MeV$ 
\cite{Nussinov3} follows from the lack of a prominent
enhancement of the $K^+ d$ scattering cross section in the relevant 
$K^+ d$ momentum interval whose size is fixed by the Fermi motion in 
the deuteron. An even stronger upper bound of $\Gamma < 1\,MeV$ is 
derived from a  phase shift analysis \cite{Arndt} and by 
analyzing along similar lines the $K^+$ charge exchange reaction 
\cite{Cahn2}. 
\\
Motivated by this new development we would like to suggest schemes 
where the width can be as small as $\Gamma < 1-3\,MeV$ and to 
dispel the belief that, when decays into pseudo scalar particles 
are allowed and the Zweig rule does not apply, the exotics are always 
extremely broad. Thus a very different 
explanation why multi-quark exotics have not been seen 
before emerges. Their widths are no longer too large but too small,  
causing the production cross sections for these exotic states scaled 
by their widths $\Gamma$ to be so tiny that the peaks in the invariant 
mass distribution escaped detection in earlier lower statistics and/or 
lower resolution experiments.
\\
A $Z^+(1530)$ state with a width of $5-15\,MeV$, $I=0$ and 
$J^\pi = \frac{1}{2}^+$ has been predicted in an extended SU(3)-flavor 
Skyrme model \cite{DPP,Polyakov}. It appears as the isospin 
singlet tip of an anti-decuplet where the corresponding 
$I=\frac{1}{2}$ doublet has been identified with the $N(1710)$ 
$J^\pi = \frac{1}{2}^+$ state. Diakonov, Petrov and Polyakov also 
suggest that the $\Sigma(1890)$ and $\Xi(2070)$ states are the 
remaining penta-quark states of this anti-decuplet, leaving only 
the $Z^+$ to be discovered.
\\
With the $Z^+(1530)$ and the corresponding measured $\Theta^+(1540)$ 
at hand we should anchor the penta-quark scale at $1540\,MeV$ and look 
for other even lighter penta-quark states comprising $u,d$ and $s$ quarks 
and anti-quarks.
\\
In the anti-decuplet the $S=0$ ~$N(1710)$ is obtained from the $Z^+$ 
by the U spin lowering operator, that replaces in 
$Z^+ \equiv \bar{s}uudd$ either one $d$ by a $s$ or one $\bar{s}$ by 
a $\bar{d}$, yielding 
\begin{equation}
\label{e.z+}
|N\,,\,I=\frac{1}{2}\,,\,I_z =-\frac{1}{2}> = 
\sqrt{\frac{1}{3}}|\bar{d}uudd> + \,\sqrt{\frac{2}{3}}|\bar{s}uuds>.
\end{equation}
The matrix element of the SU(3) breaking Lagrangian 
$m_{s}\bar{\Psi}_s\Psi_s-m_{d}\bar{\Psi}_d\Psi_d$ in the $N(1710)$ state is 
$\frac{4}{3} \Delta_m^{penta}$ and $\Delta_m^{penta}$ in the 
case of $Z^+(1530)$. $\Delta_m^{penta}$ is an effective $s$ 
and $d$ quark mass difference 
which subsumes also the hyperfine mass splitting effects.
The mass differences between these $I=0$ and $I=\frac{1}{2}$ 
penta-quark states is then
\begin{equation}
\label{e.dmnp}
m(N(1710))-m(Z^+(1530)) = 
\frac{4}{3} \Delta_m^{penta} - \Delta_m^{penta} =
\frac{1}{3} \Delta_m^{penta} = 180\,MeV.
\end{equation}
This large effective strange versus up or down quark mass difference 
in the penta-quark system $\Delta_m^{penta}=540\,MeV$ is more than 
three times larger than the difference in the standard baryonic 
decuplet $\Delta_m^{baryon}=160\,MeV$ that is traditionally 
identified with the constituent mass difference between the 
strange $s$ and non-strange $u,d$ quarks. As we indicate next 
a large $\Delta_m^{penta}=200-400\,MeV$ could be a better 
prediction reflecting the large hyperfine splittings in case 
of the $q_i\bar{q}_j$ Nambu-Goldstone pions associated with the 
spontaneous chiral symmetry breaking which the Skyrme model 
incorporates. 
\\
General arguments based on QCD inequalities motivate 
\begin{equation}
\label{e.dmx}
\Delta_m^{x} = m(X\bar{s}) - m(X\bar{q}) > m_s^{c}-m_q^{c} = \Delta_m^{c} .
\end{equation}
This conjecture constituting a stronger variant of 
Vaffa-Witten's rigorous result \cite{Vafa} will be addressed separately 
in detail. The masses on the left are those of the lowest 
lying states with given $J^\pi$ consisting 
of the same subsystem $X$ with an extra $\bar{s}$ or $\bar{q}$ 
quark in the same overall state and $m_{q,s}^{c}$ are the current 
quark masses. $\Delta_m^{c}$ is smaller than the standard 
$\Delta_m^{baryon}$ of $160\,MeV$ in common quark models.
It has been noted \cite{Shrock,Karliner} 
that $\Delta_m = m(Q\bar{s}) - m(Q\bar{q})$ monotonically decreases 
with $m_Q$. This is expected from the $1/m_Q$ decrease of the 
attractive hyperfine interaction.
\begin{equation}
\label{e.hfi}
<\Psi_H|{\cal{L}}_{hfi}|\Psi_H>= \sum_{i,j} C_{i,j}(H) \cdot 
\frac{({\vec{\sigma}_i} \cdot {\vec{\sigma}_j})({\vec{\lambda}_i} 
\cdot {\vec{\lambda}_j})}{m_{q_i} \cdot m_{q_j}}
\end{equation}
The sum extends over all the quarks and anti-quarks in the hadron $H$ 
and $C_{i,j}(H)$ is the relative wave-functions at zero separation 
of the various $q_i q_j$ or 
$q_i\bar{q}_j$ pairs in the hadron $H$. The components of $\vec{\sigma}$ 
are the $2\times2$ Pauli matrices $\sigma_{\alpha}$ with 
$\alpha \in \{1...3\}$ and those  
of $\vec{\lambda}$ are the $3\times3$ Gelman 
matrices $\lambda_{\beta}$ with $\beta \in \{1...8\}$ representing 
the eight SU(3) color generators in the ${\bf{3}}$ and ${\bf{\bar{3}}}$ 
basis of 
quarks and anti-quarks. For pseudo-scalar hadrons this differences 
$\Delta_m = m(Q\bar{s}) - m(Q\bar{q})$ starting for light quarks 
with $m(K) - m(\pi) \approx 360\,MeV$ decrease all the way down 
to $m(B_s) - m(B_u) \approx 90\,MeV$ for $b$ quarks. Nussinov and 
Shrock have conjectured \cite{Shrock} that indeed the latter mass 
difference more correctly represents the current quark mass difference 
$\Delta_m^{c}$ in agreement with lattice results \cite{Manohar}.
One has to correct for running quark masses \cite{Witten2} 
since, due to the increased reduced mass, the $Q\bar{q}$ state 
is smaller than the corresponding $q\bar{q}$ state. 
The empirical strange non-strange $\Delta_m^{baryon}$ values 
of $150-250\,MeV$ can now be explained by this smaller $\Delta_m^{c}$ 
and hyperfine interactions using constituent quark masses similar 
to those in the mesons and the fact that the hyperfine splitting 
in baryons is reduced relative to that in mesons due to 
$<baryon|{\vec{\lambda}_i} \cdot {\vec{\lambda}_j}|baryon> = 
1/2 <meson|{\vec{\lambda}_i} \cdot {\vec{\bar{\lambda}}_j}|meson>$.
\\
Next we would like to identify the likely 
non-strange analogue of the $\Theta^+ (1540)$ penta-quark and
estimate in the same manner $\Delta_m^{penta}$.
\\
In many multiplets like the vector and tensor meson nonets the 
physical mass eigenstates are adequately represented by the $I=0$
non-strange $|(u\bar{u}+d\bar{d})/\sqrt{2}>$ and strange 
$|s\bar{s}>$ quark states and are far from the SU(3) flavor singlet
$|(u\bar{u}+d\bar{d}+s\bar{s})/\sqrt{3}>$ and 
$|(u\bar{u}+d\bar{d}-2s\bar{s})/\sqrt{6}>$ octet state.
This is due to the fact, manifested in the Zweig rule, that the 
$s\bar{s} \rightleftharpoons gluons \rightleftharpoons u\bar{u}$
off diagonal matrix elements are smaller than the mass difference 
between the two strange and two non-strange ($u$ or $d$) quarks. 
\\
An important exception to this are the mesons $\pi$, $K$, $\eta$ 
and $\eta'$ of the pseudo-scalar octet with a strong coupling to 
the pure glue channel, often attributed to the U(1) axial anomaly. 
The Skyrme model contains the U(1) axial anomaly and the Skyrmions can 
be viewed as a coherent $\pi,K,\eta$ fields. Hence the emergence 
in first order of the SU(3) symmetric anti-decuplet of Diakonov, 
Petrov and Polyakov can be understood \cite{Gudkov}. Note however 
that a detailed analysis of the radiative widths \cite{Amsler} 
suggests that also the $\eta$ and $\eta'$ are intermediate states 
between the mixed and pure SU(3) flavor states.
\\
We explore the possibility that the lowest lying 
partners of the $\Theta^+(1540)$ are essentially the pure non-strange 
$\bar{u}(ud)(ud)$ and $\bar{d}(ud)(ud)$ penta-quark states 
$P^0$ and $P^+$.  
\\
In the penta-quark state we 
encounter for the first time two di-quarks taken in the 
simplest model \cite{Jaffe3,Nussinov3} to be the standard strongly 
bound $I=0$ ~$S=0$ $(u_1d_1)$ and $(u_2d_2)$ with an relative 
angular momentum $L_{12} = 1$. 
The task of estimating the difference between the analog 
states $m(\Theta^+)-m(P)$ where $P$ is obtained  by exchanging 
$\bar{s}$ in $\Theta^+ \equiv \bar{s}(u_1d_1)(u_2d_2)$ by $\bar{u}$ 
or $\bar{d}$ seems easier,
\begin{equation}
\label{e.dmtp}
m(\Theta^+)-m(P) = \Delta_m^{c} - 
\sum_{q \in \{u,d\}, i \in \{1,2\}}\!\!\!\!\!\!
C_{\bar{s},q_i} \cdot 
\frac{({\vec{\sigma}} \cdot {\vec{\sigma}_i})({\vec{\bar{\lambda}}} 
\cdot {\vec{\lambda}_i})}{m_{s} \cdot m_{q}} +
\sum_{q \in \{u,d\}, i \in \{1,2\}}\!\!\!\!\!\! 
C_{\bar{q},q_i} \cdot 
\frac{({\vec{\sigma}} \cdot {\vec{\sigma}_i})({\vec{\bar{\lambda}}} 
\cdot {\vec{\lambda}_i})}{m_{q}^2}
\end{equation}
where we sum only over the hyperfine interactions between the 
anti-quark and the four quarks in the penta-quark state because 
the remaining mutual $q_iq_j$ interactions cancels out. 
Since each of the di-quarks $u_id_i$ has $S=0$ the 
expectation value of each of the four hyperfine interactions 
vanishes. In this lowest order approximation we thus find 
$m(\Theta^+)-m(P) \approx \Delta_m^{c} \approx 100\,MeV$.
But this result has to be questioned because the penta-quark state 
with two rigid $I=0$ ~$S=0$ $ud$ di-quarks is only justified if
the remaining anti-quark is a heavy $c$ or $b$ quark with 
small mutual $\bar{Q}q_i$ hyperfine interactions. 
This is no longer the case when the anti-quark is a $\bar{u}$ or 
$\bar{d}$ because lower energies are obtainable if a $S=1$ 
admixture in the di-quark is allowed while benefiting from the 
very large $\bar{q}q$ hyperfine interaction with $S_{\bar{q}q}=0$. 
The scale for the strength of this hyperfine interaction is set 
by $m(K_{\bar{s}q})-m(\pi_{\bar{q}q}) \approx 360\,MeV$. This as 
well as the large mass splitting in the Diakonov, Petrov and 
Polyakov anti-decuplet widens the likely $m(\Theta^+)-m(P)$ range 
from $100\,MeV$ up to $350\,MeV$.
\\
Such a non-strange penta-quark $P$, with a mass of $1200-1450\,MeV$ 
and $I=1/2$ could be seen in $\pi^- p$ but not in $\pi^+ p$ system.
A new resonance cannot be ruled out by $\pi^- p$ phase shift analyzes, 
if the resonance is as narrow as $\Gamma < 5\,MeV$ and its cross 
section small~\footnote{The total formation cross section at the 
maximum of the $P$ resonance in the $\pi^- p$ channel of $5-25\,mb$ 
is smaller than the one in the $K^+ n$ channel at 
$W(m_{\Theta^+})$ of 37 mb \cite{Nussinov3}.}. 
The full circle in the Argand diagram is completed within an interval 
$\Delta_W \approx 2\Gamma$ requiring a high energy resolution and a 
small step width of less than $0.5\,MeV$ to measure the Argand diagram, 
not taking into account that even in $0.5\,MeV$ $W$ bins the total 
$P$ to $\pi^- p$ cross section ratio is only $0.1-0.3$.  
In one of the few high resolution experiments narrow peaks in 
$\pi^- p$ channel have indeed been seen in the reaction 
$^{12}C(e,e'p\pi^-)^{11}C$ at MAMI \cite{MAMI}. The high missing 
mass resolution of $\sigma_m = 0.27\,MeV$ allows to identify 
two $4\,MeV$ narrow states at an invariant mass of $1222\,MeV$ 
and $1236\,MeV$ that have been interpreted as bound $\Delta^0$ states. 
Interestingly a narrow resonance with a mass of $1225\,MeV$ and a 
width of $50\,keV$ was found to diminish the $\chi ^2$ of partial wave 
analyzes based on the $\pi N$ SAID data base \cite{Azimov}.
\\
The most puzzling aspect of the $\Theta^+(1540)$ is the narrow width 
$\Gamma < {\cal{O}}(1-3\,MeV)$ established by $K^+$ nuclear scattering 
data. If the $\Theta^+(1540)$ will survive further experimental scrutiny 
and even more so if lighter narrow analogues exist, then finding a 
credible scenario for such narrow widths is a challenge to QCD, 
which should at least qualitatively explain all hadronic data.
The penta- and tetra-quark are more complex than ordinary baryons and 
mesons. In a chromoelectric flux tube model (CFTM) mesons are tubes  
with the radius b of chromoelectric flux extending between the quark 
and anti-quark at both ends. It generates a confining linear potential
$V = \sigma \cdot r_{q\bar{q}}$ where $\sigma$ is the string tension 
\cite{CNN,Isgur}. This picture is 
particularly appropriate for states with large angular momenta and length 
$r_{q\bar{q}} = a \gg b$. In this limit linear Regge trajectories with 
a slope $\alpha' = 2\pi\sigma \approx 1\,GeV^2$ arise and in the 
limit of $b \rightarrow 0$ a string like model emerges.
\\
In the CFTM baryons consist of three flux tubes which join at a junction
$J$. The tetra- and penta-quarks have more elaborate color networks
with two and three junctions respectively (see Fig.\ref{pic.tube}). 
The complexity of the penta-quark state suppresses its formation
in hadronic collisions due to the specific rearrangements that are
required.  By detailed balance the system will then also be trapped
for a long time while finding the specific path back to the original
hadrons. The CFTM  offers a concrete realization of this general
concept. To allow for the penta-quark state to decay into baryon and
meson or for the tetra-quark state to decay into two mesons a junction $J$ 
and the anti-junction $\bar{J}$ have to annihilate before the flux tubes 
can reconnect to generate the final two hadrons.
A necessary but not sufficient condition for the annihilation is that 
$J$ and $\bar{J}$ are within a relative distance smaller than the flux 
tube radius $b$.
In a simplistic terta-quark model we identify the centers of mass of the
di-quarks $q_1q_2$ and $\bar{q'}_1\bar{q'}_2$ with the junctions $J$ 
and $\bar{J}$. These junctions are treated as scalar particles with 
the mass of the di-quarks $m_{dq} \approx 650\,MeV$. These particles 
are in the color representation ${\bf{\bar{3}}}$ and ${\bf{3}}$ with the 
same linear 
potential as between the quarks in a meson. The two junctions are in a  
relative $p$-wave. Let $<r_{J\bar{J}}>$ in the lowest state be $a$. 
Since the $l=1$ wave-function decreases linearly in the small $r$ region, 
the probability $P(r < b)$ is approximately $(b/a)^5$. For 
$\sigma \approx 0.15\,GeV^2$ we estimate, using the minimum of the radial 
potential and the curvature $d^2V/dr^2$ at this minimum, that 
$r_{min} \approx \;<\!\!r\!\!>\; \approx a$ is larger than $0.7\,fm$. 
For $b$ in the range of 
$0.2-0.5\,fm$ the probability $P(r<b)$ is smaller than $0.002-0.2$.
\\
To estimate the decay rate note that the relative coordinate $r$ changes 
by $2a$ during a period $T$ exceeding $a/c \approx 0.7\,fm/c \approx 
(300\,MeV)^{-1}$. During each period $J\bar{J}$ annihilation occures 
with a probability which is smaller than $P(r<b)$. Hence the decay rate is  
$\Gamma^{Tetra} < P(r<b) \cdot 300\,MeV = 0.6-60\,MeV$. In the penta-quark 
either of the two junctions $J$ (see Fig.\ref{pic.tube}) can annihilate 
with $\bar{J}$, therefore we expect roughly twice this width.
\\
The crucial flux tube radius $b$ can be determined by pure gluo-dynamics 
up to small $1/N_c^2$ corrections. The large mass of the lightest 
glue-ball is expected to be above $m_{gb} \ge 1.6\,GeV$. This sets the 
scale for the radius of the flux tube to $b = 1/m_{gb} \le 0.15\,fm$. 
A full QCD lattice simulation, recently performed for $QQQ$ baryons with
three quarks pinned down at relative distances of 
$r_{qq} \approx 0.7\,fm$, clearly reproduces the $Y$ shape of the flux 
structure shown in an action density contour plot \cite{Ichie}. 
The corresponding flux junction radius is $b \le 0.2\,fm$.
Even in the ground state $qqq$ nucleons, where the 
light quarks move so fast that the short flux tubes get tangled up 
into a more uniform spherical distribution than in the case of the 
$\bar{q}(ud)(ud)$ penta-quark states, the flux junction radius will 
remain the same.
\\
On the other hand the distance $a$ between $J$ and $\bar{J}$ in 
tetra-quarks is
related to the distance between the quark and anti-quark in mesons, 
because both have the same color charges and the same flux tubes.
In mesons $a$ is essentially the charge radius of the pion 
$r_{\pi} \approx 0.65\,fm$. The di-quark masses are $m_{dq}=650\,MeV$ 
instead of $350\,MeV$ for the light quarks. For a linear potential 
this scales the size by $(m_{q}/m_{dq})^\frac{1}{3}$. However here, 
the two di-quarks are in a 
$p$-wave. The resulting additional centrifugal barrier effectively 
doubles the kinetic energy reducing the mass by a factor of about $2$, 
which restores $<r_{J\bar{J}}> = a$ back to out initial estimate of 
about $0.7\,fm$.
\\[5mm]
\noindent
{\large \bf{Acknowledgments}} \\[3mm]
We like to thank A. Casher for many illuminating discussions of the 
flux tube model.
\\[-10mm]

\renewcommand{\refname}{}

\end{document}